\documentclass[reprint, 10pt, a4paper, aps, prx, floatfix]{revtex4-1}
\usepackage[utf8]{inputenc}
\usepackage{lipsum}
\usepackage{graphicx}
\usepackage{dcolumn}
\usepackage{bm}
\usepackage{amssymb}
\usepackage{amsmath}
\usepackage{textcomp}
\usepackage{color}
\usepackage{mathtools}
\usepackage{units}

\begin{document}

\title{Focusability in the multi-pump Raman amplification of short laser pulses}
\author{K. V. Lezhnin$^{1,2}$}
\email{klezhnin@pppl.gov}
\author{K. Qu$^{2}$, N. J. Fisch$^{2}$}
\affiliation{{$^{1}$Princeton Plasma Physics Laboratory, 100 Stellarator Rd, Princeton, NJ 08540, USA}}
\affiliation{{$^{2}$Department of Astrophysical Sciences, Princeton University, Princeton, New Jersey 08544, USA}}

\date{\today}

\begin{abstract}
Spatially combining multiple strong laser beams is a promising concept for achieving ultrahigh laser intensities. Proof-of-principle experiments have been conducted at the National Ignition Facility to report a combination of up to twenty pulses with high energy conversion efficiency. However, the combination process might damage the seed focusability due to mismatch of the seed and pump wavefronts. Here, we investigate the effect of the finite pump beam size on the focusability of the seed pulse. We propose an approach to retain and even improve the seed focusability by specifically arranging multiple pump beams. The results are demonstrated by numerical solution of coupled nonlinear Schr\"{o}dinger equations. 

\bigskip

\end{abstract}

\maketitle

\section{Introduction}\label{intro}

The development of high-power laser technology has nearly reached the stage that any further improvement requires combining multiple laser beams into one beam in plasma \cite{Danson2019}. The beam combination process overcomes the issues of thermal damage and other deleterious nonlinear effects of a conventional amplifying medium. Combining separately amplified beams in the spectral domain, known as chirped pulse amplification (CPA) \cite{CPA}, has become the state-of-the-art method of producing high laser intensities. Beam combination can also take place in the real domain provided that nonlinear resonant processes in plasmas, such as Raman scattering~\cite{Malkin1998} or Brillouin scattering~\cite{Andreev2006}, efficiently transfer laser energy among different pulses~\cite{Trines2020}. This idea was already demonstrated with high pump-to-seed conversion efficiency at the National Ignition Facility (NIF)~\cite{Kirkwood2018}. While the experiments were carried out with terawatt-scale laser beams, their results suggest scalability up to hundreds of terawatt scale, broadening the applicability of beam combination for large laser facilities like NIF and OMEGA.

To achieve high energy conversion efficiency, it is assumed that the seed pulse of the combiner has a larger spot size than the pump beams when they interact. The amplified seed pulse is then tightly focused to reach the peak intensity. However, the concern is that the mismatch of the seed and pump profiles may irreversibly distort the seed pulse wavefront, thereby limiting its focusability. It was shown in \cite{Jia2020} that such phase distortion takes place in the linear regime of Raman amplification because the trailing plasma wave continuously accumulates phase and transfers it to the amplified pulse. It was also demonstrated that the phase distortion may be mitigated by working in the nonlinear regime to eliminate lagging between the seed and the amplified pulse. It, however, assumes a plane wave pump for homogeneous amplification, which cannot be satisfied in beam combination. 

In the present work, to analyze how the finite pump beam spot size impacts the seed focusability, we assume that the seed and pump interact via backward or side Raman scattering in plasmas. We focus on the strong-seed nonlinear regime which is the most efficient for energy transfer. In this regime, the pump, when interacting with the seed, completely transfers its energy into the seed while maintaining the seed pulse phase. In effect, each pump pulse locally modulates the seed amplitude without changing its phase. Denoting the unamplified seed profile by $b_0(\bm{r},z_0)$ and the amplitude modulation function by $M(\bm{r})=|a(\bm{r})|\delta(z-z_0)$ (which is solely determined by the pump pulse profile), the amplified seed profile is then $b(\bm{r},z_0) = b_0(\bm{r},z_0)M(\bm{r})$. Using the convolution formulation of Fresnel diffraction \cite{hecht2017}, evolution of the amplified pulse is described as $b(\bm{r},z) = b(\bm{r},z_0)*h(\bm{r},z)$ where $*$ denotes convolution over the $\bm{r}$ plane and $h(\bm{r},z)=1/(i\lambda z) e^{-ikz} e^{-ik \bm{r}^2/(2z)} $ is the impulse response function. The formulation thus separates the diffraction of the unamplified seed and the modulation function, $b(\bm{r},z) = [b_0(\bm{r},z_0)*h(\bm{r},z)]\cdot m(\bm{r},z) = b_0(\bm{r},z)m(\bm{r},z)$ with $m=M(\bm{r})*h(\bm{r},z)$. We now see that the amplified seed propagates identically to the unamplified seed and then gets amplitude modulated by the function $m(\bm{r},z)$ which itself obeys Fresnel diffraction. Although the modulation function at the interaction plane $M(\bm{r})$ has a flat phase, it gradually develops a $\bm{r}$-dependent phase from Fresnel diffraction at $z>z_0$ and would cause distortion to the seed phase front in its focal plane. 
As we will show below, retaining the seed focusability requires that $m(\bm{r},z)$ has a flat phase within the seed spot size, i.e. $M(\bm{r})$ is sufficiently large. 

{While it may be challenging to arrange a single wide pump pulse from the experimental standpoint, the same result could be achieved by utilizing multiple pump beams that will cover the full seed beam. The pumps, however, need to interact with the seed in the same plane, otherwise each would diffract before the adjacent beams are able to compensate for the phase shift. More interestingly, one can create a modulation function by design such that $m(\bm{r},z)$ is highly concentrated within the central parts of the seed, 
thereby the amplified seed can focus to higher intensity than the Gaussian pulse with the energy of Raman-amplified seed. We will show in Sec.~\ref{results} that such a modulation function can be produced by using the beat wave of the pumps.}

Raman amplification using multi-pump configurations has been investigated in \cite{Fraiman2002,Solodov2003,Balakin2003,Edwards2017}. The pump pulses are considered incoherent or partially coherent which is treated by combinations of different frequency components. These studies demonstrate the resilience of Raman amplification to pump incoherence or large scale plasma density fluctuations. Our current paper considers speckles of the pump and describes how they affect the prefocused seed.

The paper is organized as follows. Section \ref{theory} contains a simple theory of amplified seed pulse focusability that will be used for the interpretation of the numerical results. We also explain how the amplification only modulates the seed amplitude but not the phase front. In Section \ref{setup}, we formulate the numerical model using coupled wave equations with each laser pulse modeled by nonlinear Schr\"{o}dinger equation (NSE). In Section \ref{results}, we present a case study of different pump configurations, showing numerical solutions of the NSE and comparing them with analytical estimates. Finally, we summarize and discuss our results in Section \ref{discussion}.

\section{Simple theory of the amplified seed focusability}\label{theory}

\begin{figure}
    \centering
    \includegraphics[width=\linewidth]{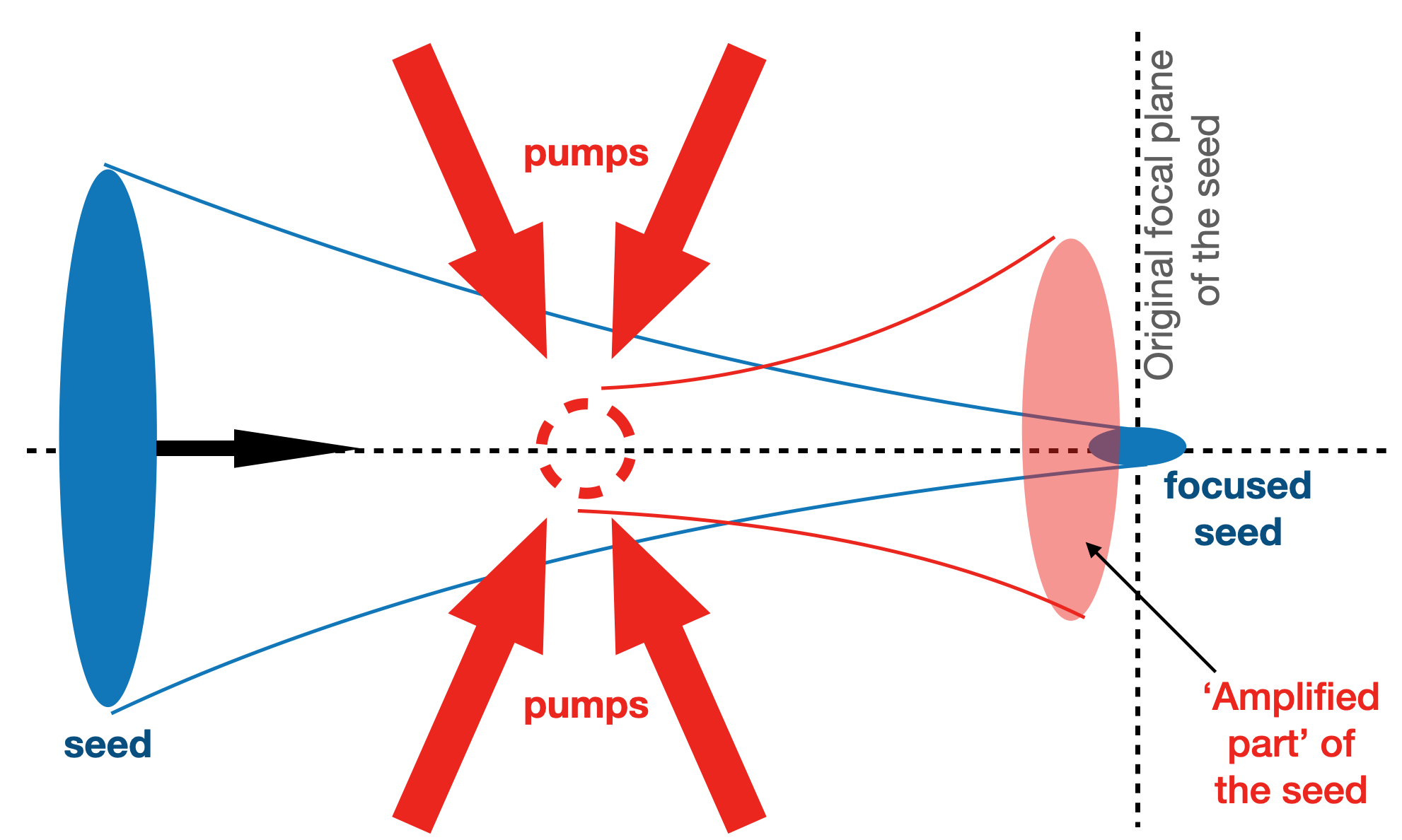}
    \caption{Setup of the problem. Seed pulse is tightly focused and is amplified by the pump pulses of fixed total energy far away from the seed focal plane, $|z_{\rm ampl}-z_{\rm foc}|\gg z_R$. The objective is to optimize the amplified seed intensity at the original focal plane of the seed.}
    \label{fig:setup}
\end{figure}

In this section, we expand upon the analytical description of the amplified seed focusability that was briefly mentioned in the previous section. Suppose a Gaussian seed pulse which propagates along the $z$ axis from $z=-z_0$ and focuses at $z=0$ plane. It may be written down using $q$-parameter as follows:

\begin{equation}
    b_0({\bf r},z) = \frac{1}{q(z)} \exp{\left[\frac{-ik{\bf r}^2}{2q(z)}\right]},
    \label{eqn:seed}
\end{equation}

\noindent where $q(z)=z+i z_R$, with $i$ denoting complex unity and $z_R=N_{\rm refr} \pi w_0^2/\lambda$ being seed pulse Rayleigh length, $N_{\rm refr}=(1-\omega_{\rm pe}^2/\omega_0^2)^{1/2}$ - refractive index of plasma, $w_0$ - seed pulse waist, $\lambda$ - seed wavelength, $k=N_{\rm refr}2\pi/
\lambda$. Assume we were able to amplify the seed by an array of pump pulses in a instantenous and local manner (see Figure~\ref{fig:setup} for the overall setup of the problem). This approximation may be valid, for instance, in the deeply nonlinear regime of Backward Raman Amplification \cite{Malkin1998,Trines2020}. We may write down the profile of the corresponding pump pulse as follows:

\begin{equation}
    a({\bf r},z)= \delta{(z-z_0)}\sum_{i=1}^{\rm N_{\rm pump}}a_{0,i} \exp{\left[-\frac{({\bf r}-{\bf r}_{a,i})^2}{w_{a,i}^2} \right]}.
    \label{eqn:pump}
\end{equation}

\noindent Here, $a_{0,i}$ is the $i$-th pump amplitude at the amplification plane, 
${\bf r}_{a,i}$ is the radius-vector of the axis of symmetry of the $i$-th pump pulse, $w_{a,i}$ is the pump width, and $z_0<0$ denotes the coordinate where the pump energy is deposited to the seed pulse. We also require the total pump energy to be fixed regardless of the pump configuration:

\begin{equation}
    \int dV |a({\bf r},z)|^2 = \rm const.
    \label{eqn:pumpenergy}
\end{equation}

\noindent The resulting amplified seed envelope at the amplification plane, $b({\bf r},z_0)$, may be expressed as $b({\bf r},z_0)=b_0({\bf r},z_0)(1+a({\bf r},z_0))$. Here, $a({\bf r},z_0)$ is a real number representing the amplitude profile of the pump pulse. The underlying assumption is that the phase of the amplified seed is not affected by the Raman amplification process. This is because the pump-to-seed energy scattering is mediated by the plasma wave. The phase front of the plasma wave is determined by the beat of pump and seed, so the scattered wave from the pump matches the phase fronts of the seed. (The opposite is also true, although anti-Stokes scattering from the seed is much weaker.) In Backward Raman Amplification \cite{Malkin1998}, the nearly stationary plasma wave travels towards the seed tail, which could cause phase mixing. But since the seed pulse is self-compressed towards its front, its tail becomes negligible asymptotically, as demonstrated in \cite{Fraiman2002}. In the near-forward Raman scattering and side Raman scattering, the plasma wave only interacts with the laser where the pump and seed overlap, so this assumption is exact.

Denoting the amplified part of the seed as $b_a({\bf r},z_0) \equiv b_0({\bf r},z_0) a({\bf r},z_0)$, we may calculate the evolution of the amplified seed pulse envelope as seed propagates using the convolution formulation of Fresnel diffraction:

\begin{align}
    &b_a({\bf r},z) = b_a({\bf r},z_0) \ast \frac{1}{i\lambda z}e^{-ikz} \exp{\left[-\frac{ik{\bf r}^2}{2z} \right]} \nonumber \\ 
    &=\int d^2 {\bf r'} b_a({\bf r'},z_0) \frac{e^{-ik(z-z_0)}}{i\lambda (z-z_0)}\exp{\left[-\frac{ik({\bf r-r'})^2}{2(z-z_0)} \right]}
    \label{eqn:Fresnelintegral}
\end{align}

\noindent By calculating the value of such convolution, one may find the transverse profile of the amplified seed pulse at the focal plane of the original seed, $|b_0({{\bf r},0})+b_a({{\bf r},0})|^2$. 
Calculating such metric and comparing its values for various pump configurations $(a_{0,i},w_{a,i},{\bf r}_{a,i})_{i=1}^{N_{\rm pump}}$, we may analyze the effect of the pump configuration space on the seed focusability. In what follows, first we will consider a few illustrative configurations that will reveal the effects of the multi-pump amplification on the seed focusability and then illustrate them using NSE simulations in Section \ref{results}.

\subsection{Single Gaussian pump of finite width}

Let's start with the simple case of a single Gaussian pump pulse that symmetrically amplifies the seed (i.e. ${\bf r}_{a}=0$):

\begin{equation}
    a({\bf r},z)= a_{0} \exp{\left[-\frac{{\bf r}^2}{w_{a}^2} \right]} \delta{(z-z_0)}.
\end{equation}

\noindent Calculating the amplified part of the seed from Eqn.~\eqref{eqn:Fresnelintegral}, we obtain the following:

\begin{equation}
    b_a({\bf r},z) = -a_0\frac{e^{-ik(z-z_0)}}{q_{\rm new,a}} \exp{\left[\frac{-ik{\bf r^2}}{2q_{\rm new,b}} \right]},
    \label{eqn:amplseedonepump}
\end{equation}

\noindent with $q_{\rm new,a}$ and $q_{\rm new,b}$ being an effective complex beam parameters of the amplified part of the seed. These two new $q$ parameters are generally distinct (hence, the amplified part of the seed is not Gaussian) and are given by the following expressions:

\begin{flalign}
    q_{\rm new,a} &= z+\frac{w_b^2}{w_a^2}(z-z_0)+i \left[z_R-\frac{w_b^2}{w_a^2}\frac{z_0}{z_R}(z-z_0) \right], \label{eqn:qnewa} &\\
    q_{\rm new,b} &= z-z_0\left[1-\frac{1}{1+2\frac{w_b^2}{w_a^2}+\frac{w_b^4}{w_a^4}(1+\frac{z_0^2}{z_R^2})}\right] \nonumber &\\ 
    &+ iz_R \left[\frac{1+\frac{w_b^2}{w_a^2}\left(1+\frac{z_0^2}{z_R^2}\right)}{1+2\frac{w_b^2}{w_a^2}+\frac{w_b^4}{w_a^4}\left(1+\frac{z_0^2}{z_R^2}\right)}\right]. \label{eqn:qnewb}
\end{flalign}



\noindent In the limit of $w_b/w_a \ll 1$, we recover $q_{\rm new,a}=q_{\rm new,b}=q(z)$, i.e. regular Gaussian envelope.

From the shape of Eqn.~\eqref{eqn:qnewa}, one may see that for a seed pulse focused at the $z=0$ plane, the focal plane of the amplified part of the seed is shifted by $-z_0 w_b^2/w_a^2$, and the diffraction rate given by $z_R (1+w_b^2/w_a^2\cdot z_0^2/z_R^2)>z_R$, meaning that the diffraction is faster for the amplified part of the seed. These results put a strong limitation on the overall focusability of the amplified seed.





\begin{figure}
    \centering
    \includegraphics[width=\linewidth]{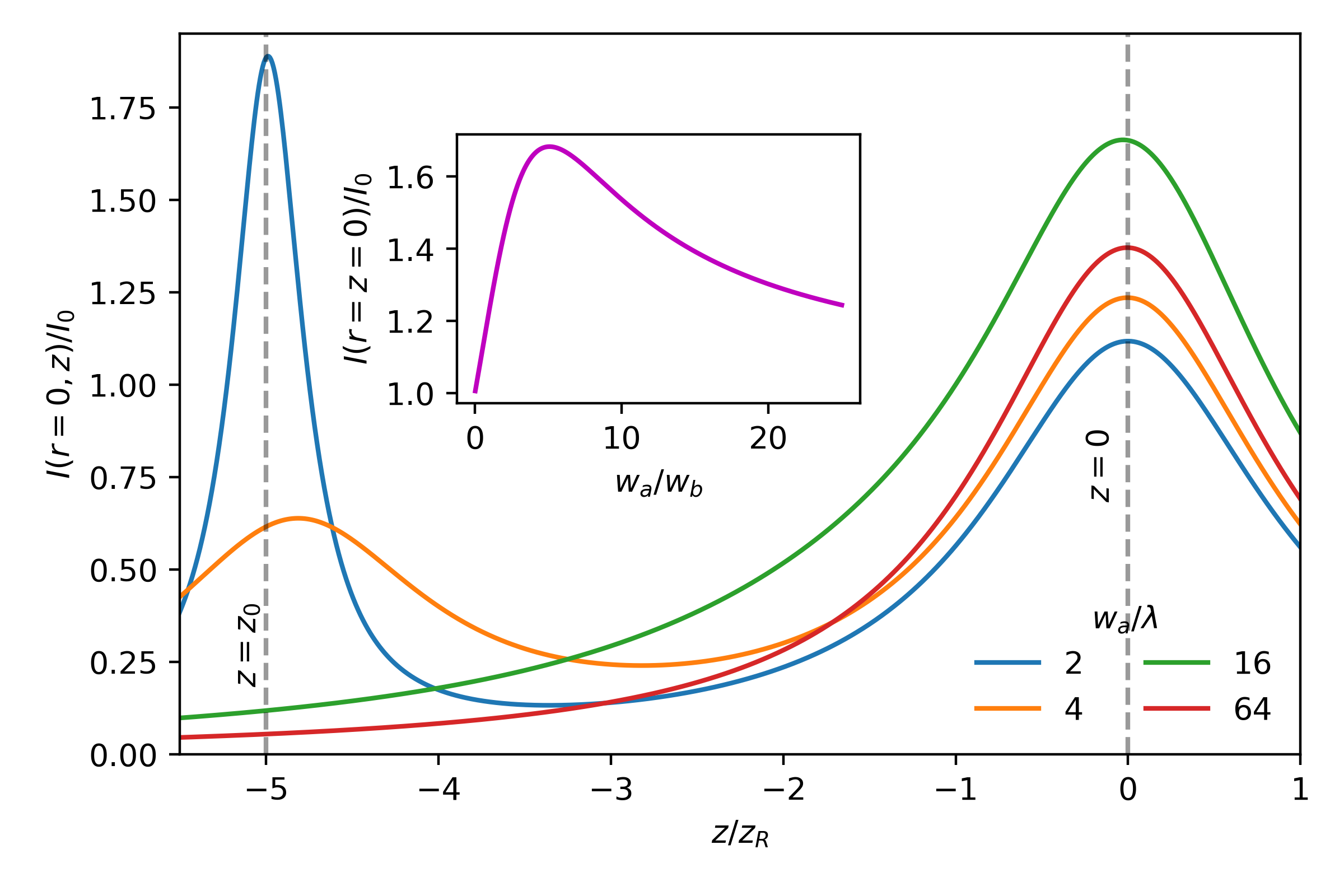}
    \caption{Intensity of the amplified seed on laser axis ($r=0$) along the propagation direction and dependence of the intensity on original focal plane ($r=0,z=0$) on pump width.}
    \label{fig:intensityvswidthvsz}
\end{figure}

Using the result given by Eqns.~\eqref{eqn:amplseedonepump}-\eqref{eqn:qnewb}, one may calculate the peak intensity at the original focal plane (i.e., at the $z=0$ plane) and the location of the peak intensity on the $z$ axis. The on-axis intensity of the amplified seed ($I_a\equiv I_a(r=0,z)=|b_0(r=0,z)+b_a(r=0,z)|^2|$) may be written as follows:

\begin{equation}
    I_a = \left| \frac{1}{z+iz_R}+\frac{a_0}{z+iz_R +(z-z_0)\frac{w_b^2}{w_a^2}\left(1-i\frac{z_0}{z_R}\right)} \right|^2.
    \label{eqn:intensityonepump}
\end{equation}


\noindent Since we aim to consider pumps at fixed power for any particular configuration of the pump parameters, it puts a restriction on the pump intensity and width: $w_a^2 a_0^2=\rm const$. Introducing seed and pump power ratio, $\alpha=P_a/P_b$, we may express $a_0 = \frac{w_b}{w_a}\sqrt{\alpha}b_0$. When the seed pulse is focused at $z=0$, the intensity may be written as:

\begin{equation}
    I_a(z=0)=\frac{1}{z_R^2}\left|i +\frac{w_a}{w_b} \frac{\sqrt{\alpha}b_0}{\frac{z_0}{z_R}-i\left(\frac{w_a^2}{w_b^2}+\frac{z_0^2}{z_R^2}\right)}\right|^2.
    \label{eqn:intensityfocalplane}
\end{equation}


\noindent From this expression, one may see the role of the pump width for the amplified seed peak intensity at the original focal plane. Figure~\ref{fig:intensityvswidthvsz} shows the evolution of the peak intensity along the propagation direction and the dependence of the focal plane intensity as a function of the pump width. It is seen that there is an optimal pump width that may be analytically derived from Eqn.~\eqref{eqn:intensityfocalplane}. For tightly focused pumps, one may see that the peak intensity is realized right after the amplification due to the high intensity of the pump pulse (see Eqn.~\eqref{eqn:Fresnelintegral}), but the focal plane intensity ends up being smaller as the amplified part of the seed diffracts. Very wide pumps deposit a significant fraction of their energy away from the central parts of the seed, which also limits the resulting peak amplitude. Thus, there is an intermediate value of the pump width where the seed focusability is optimized. The optimal width may be determined numerically using Eqn.~\eqref{eqn:intensityfocalplane}.


Another effect that may impede seed pulse focusbility is the group velocity dispersion (GVD). For fixed pump energy, a wide pump pulse corresponds to a short duration of amplified seed, which is susceptible to strong GVD. 
We next quantitatively analyze how GVD contributes to the beam quality during beam combination. Assuming a pre-chirped seed pulse is focused to the Gaussian with zero frequency chirp, we may calculate the resulting amplified seed profile at the original focal plane. Thus, we start with the one-dimensional unamplified seed:

\begin{equation}
    u(z,0) = \exp{\left(-\frac{z^2}{2L_b^2}+ik_0z\right)}
    \label{eqn:fourierintegral0}
\end{equation}

\noindent To obtain the envelope at the time of the amplification, we use the following representation of the seed envelope:

\begin{eqnarray}
    u(z,t) = \frac{1}{\sqrt{2\pi}} \int_{-\infty}^{+\infty} dk\, A(k) e^{ikz-i\omega(k)t},
    \label{eqn:fourierintegral1} \\
    A(k) = \frac{L}{2}\exp{(-0.5(k-k_0)^2L^2)},
    \label{eqn:fourierintegral2}
\end{eqnarray}

\noindent where $A=A(k)$ is the complex amplitude of the seed. Using the dispersion relation in use is $\omega(k)=\sqrt{\omega_{\rm pe}^2+k^2c^2}$ and expanding around the primary wavenumber $k_0$, we write $\omega(k) \approx \omega(k_0)+v_{\rm gr}(k-k_0)+\frac{1}{2}\frac{\partial^2 \omega}{\partial k^2}|_{k=k_0}(k-k_0)^2 \equiv \omega(k_0)+v_{\rm gr}(k-k_0)+{\kappa}(k-k_0)^2$.

\begin{figure}
    \centering
    \includegraphics[width=\linewidth]{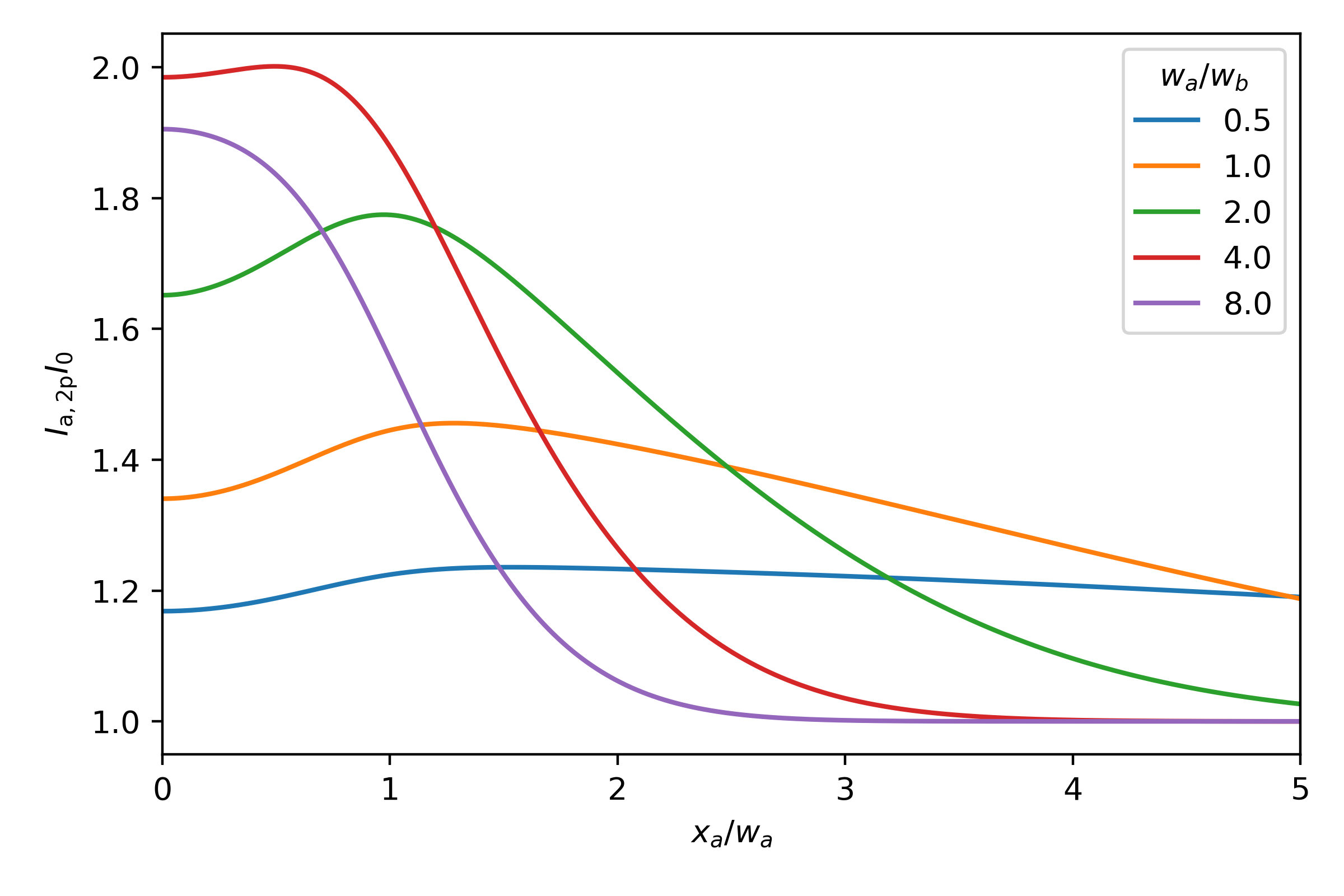}
    \caption{Peak intensities in case of two-pump amplification as the function of pump separation.}
    \label{fig:Itwo2Ione}
\end{figure}

For an arbitrary time $t$, the unamplified envelope given by integration of Eqn.~\eqref{eqn:fourierintegral1} reads:

\begin{equation}
    u(z,t) = \frac{\exp{(ik_0z-i\omega_{0}t)}}{\sqrt{1+\frac{2it\,{\kappa}}{L^2}}}\exp{\left(-\frac{(z-v_{\rm gr}t)^2}{2(L^2+2it\,{\kappa})}\right)}.
    \label{eqn:fourierintegral3}
\end{equation}

\noindent It is easy to see that for $t=0$, it reproduces Eqn.~\eqref{eqn:fourierintegral0}. Now, by calculating the amplified seed at the time of seed amplification ($t_{\rm ampl} = z_{0b}/v_{\rm gr}<0$) as $u_{\rm ampl}(z,t_{\rm ampl})=u(z,t_{\rm ampl})(1+a_0\exp{(-(z-z_{0a})^2/L_a^2)})$, we get the amplified seed at the original focal plane by following a similar procedure, which may be written as:

\begin{equation}
    u_{\rm ampl}(z,0) = IFT\left[\exp{(i\omega(k)t_{\rm ampl})}FT(u_{\rm ampl}(z,t_{\rm ampl}))\right],
    \label{eqn:fourierintegral4}
\end{equation}

\noindent where $FT$ and $IFT$ denote Fourier and inverse Fourier transforms with respect to z and k coordinates, respectively. The peak amplitude then may be compared with the peak amplitude in the case when GVD is turned off ($\kappa=0$). This results in additional pump width-dependent factor that may be added to Eqn.~\eqref{eqn:intensityfocalplane} to compare with NSE simulations.

\subsection{Two Gaussian pumps pulses}

As beam combining experiments involve multiple pump beams and since the arbitrary transverse profile may be represented as a sum of multiple Gaussian beams, let us consider the case of seed-pump amplification by two Gaussian pumps. Conducting a similar calculation using Eqns.~\eqref{eqn:pump}\&\eqref{eqn:Fresnelintegral} for the case of two pumps shifted to $x=\pm x_a, \, y=0$ axes and carrying the same total energy as in the one pump case, one may obtain the intensity on $r=0$ axis:

\begin{flalign}
    I_{a,2p} &= \frac{1}{z_R^2}\left|\frac{1}{\frac{z}{z_R}+i}+\frac{2ia_0\exp{\left[\frac{x_a^2}{w_a^2}G^{-1}(w_a/w_b,z/z_R,z_0/z_R)\right]}}{\frac{w_b^2}{w_a^2}\left(\frac{z_0}{z_R}+i\right)\left(\frac{z}{z_R}-\frac{z_0}{z_R}\right) +i\left(\frac{z}{z_R}+i\right)} \right|^2,
    \label{eqn:twopumps}
\end{flalign}

\noindent with function $G$ given by:

\begin{equation}
    G = {1+i\frac{w_a^2}{w_b^2}\frac{\frac{z}{z_R}+i}{\left(\frac{z_0}{z_R}+i\right)\left(\frac{z}{z_R}-\frac{z_0}{z_R}\right)}}.
    \label{eqn:gfunc}
\end{equation}

%


\noindent For $x_a=0$, we reproduce a single pump case given by Eqn.~\eqref{eqn:intensityfocalplane}. Figure \ref{fig:Itwo2Ione} depicts how the amplified seed intensity at the focal plane depends on the transverse shift of the two pumps. It may be seen that there is an optimal pump separation that yields highest intensity at the focal plane. To compare with NSE runs, we also incorporate the effect of GVD, analogous to the procedure outlined in Eqns.~\eqref{eqn:fourierintegral0}-\eqref{eqn:fourierintegral4}.

\section{Coupled NSE model setup}\label{setup}

Let us state the model of seed - pump interaction. First of all, we consider the seed and pumps to be in a resonance:

\begin{equation}
    \omega_{\rm p,i}=\omega_{\rm s}+\omega_{\rm pe,i}, ~
    {\bf k}_{\rm p,i}={\bf k}_{\rm s}+{\bf k}_{\rm pe,i},
\end{equation}

\noindent where $(\omega_{\rm p,i},{\bf k}_{\rm p,i})$ are the frequency and wavenumber of the $\rm i^{\rm th}$ pump, $(\omega_{\rm s},{\bf k}_{\rm s})$ - of the seed, and $(\omega_{\rm pe,i},{\bf k}_{\rm pe,i})$ are the plasma waves corresponding to Raman resonance. For the case of uniform plasma and non-relativistic seed-pump interaction, it is clear that $\omega_{\rm pe,i}= \omega_{\rm pe}={\rm const}$.

Following \cite{Lehmann2013,Li2018}, we adopt the following model for the seed-pump interaction:

\begin{flalign} 
(\frac{\partial}{\partial t} +v_{gp,i}\cdot \nabla - \frac{ic^2}{2\omega_p}\nabla_{\perp}^2- \frac{ic^2\omega_{\rm pe}^2}{2\omega_p^3}(\hat{v}_{gp}\cdot\nabla)^2 \nonumber &\\-\frac{3i\omega_{\rm pe}^2}{16\omega_p}(|\hat{b}|^2+\sum_{i=1}^N|\hat{a}_i|^2) )\hat{a}_i  &= -V\hat{b}{\hat{n}_{f,i}}, \label{eqn:NSEpump} &\\
(\frac{\partial}{\partial t}+v_{gs}\cdot \nabla - \frac{ic^2}{2\omega_s}\nabla_{\perp}^2- \frac{ic^2 \omega_{\rm pe}^2}{2\omega_s^3}(\hat{v}_{gs}\cdot\nabla)^2 \nonumber &\\-\frac{3i\omega_{\rm pe}^2}{16\omega_s}(|\hat{b}|^2+\sum_{i=1}^N|\hat{a}_i|^2))\hat{b} 
&= V \sum_{i=1}^N {\hat{n}_{f,i}^\ast} \hat{a}_i. \label{eqn:NSEseed}
\end{flalign}
\begin{eqnarray}
\frac{\partial}{\partial t}\hat{n}_{f,i} = W_{i} \hat{a}^\ast  \hat{b}_i,  \\
V = \frac{\omega_{\rm pe}^2}{4 \omega_s},  \\
 W_{i}= \frac{c^2|k_s-k_{p,i}|^2}{\omega_{\rm pe}}.
\end{eqnarray}

\noindent Here, $\hat{a_{i}}$, $\hat{b}$, and $\hat{n_{f,i}}$ are envelope functions for $i$-th pump, seed, and $i$-th pump-seed density beating, respectively. These equations include the pulse propagation terms (first and second on the LHS of Eqns.~\eqref{eqn:NSEpump},\eqref{eqn:NSEseed}), diffraction (third), group velocity dispersion (GVD) term (fourth), and the last one corresponds to relativistic self- and mutual focusing. Seed and pumps interact via the RHS of Eqns.~\eqref{eqn:NSEpump}\&\eqref{eqn:NSEseed}.

For the purpose of beam combination, the pump beams are prepared below the relativistic intensity to avoid any deleterious relativistic effects, such as self-focusing. We also introduce the normalization for the numerical calculations: $t \rightarrow \omega_s \tau$, $\nabla \rightarrow c/\omega_s \nabla$, $v \rightarrow v/c$, $V,W \rightarrow V/\omega_s,W/\omega_s$ and redefine $\hat{n}_{f,i} \equiv V \hat{n}_{f,i}$, which yields:

\begin{flalign}
\left[\frac{\partial}{\partial \tau}+v_{gp,i}\cdot \nabla - \frac{i\omega_s}{2\omega_p}\nabla_\perp^2 - \frac{i\omega_{\rm pe}^2}{2\omega_s^2}\frac{\omega_s^3}{\omega_p^3}\nabla_\parallel^2\right]\hat{a}_i& = -\hat{b}{\hat{n}_{f,i}}, \label{eqn:NSEnorm1}&\\
\left[\frac{\partial}{\partial \tau}+v_{gs}\cdot \nabla - \frac{i}{2}\nabla_\perp^2 - \frac{i\omega_{\rm pe}^2}{2\omega_s^2}\nabla_\parallel^2 \right]\hat{b} &= \sum_{i=1}^N \hat{a}_i {\hat{n}_{f,i}^\ast}, \label{eqn:NSEnorm2}\\
\frac{\partial}{\partial \tau}\hat{n}_{f,i} &= VW_{i}\hat{a}^\ast \hat{b}_i,  \label{eqn:NSEnormDens}&\\
V &= \frac{\omega_{\rm pe}^2}{4 \omega_s^2},  &\\
 W_{i}&= \frac{c^2|k_s-k_{p,i}|^2}{\omega_{\rm pe}\omega_s} \label{eqn:NSEnorm3}.
\end{flalign}

\noindent The system of Eqns.~\eqref{eqn:NSEnorm1}-\eqref{eqn:NSEnorm3} is then solved on a two-dimensional grid using symmetrized split-step Fourier approach \cite{agrawal2006} and transfer matrix method to account for seed-pump interaction via RHS terms and implemented in a Python solver \cite{CNSE}. This is a system of equations that possesses Forward Raman Scattering and Backward Raman Scattering solutions, which were numerically reproduced and in agreement with analytical solutions \cite{Qu2017}.

Let us describe the overall setup of our NSE calculations. We consider a two-dimensional box with the size of $(-200,200) \times (-150,150)$ and grid size of $512\times 512$. The seed pulse is defined as follows. Initially located at $x=-150,~y=0$, it is focused to the point $x=+150,~y=0$, with Gaussian shape at the focus having width of $4$ and duration of $10$ ($1/e$). It travels along the +x direction with the group velocity, $v_{gs}$, corresponding to the group velocity of electromagnetic wave of frequency $\omega_s$ travelling in plasma with plasma frequency $\omega_{\rm pe}$ ($v_{gs}= c\sqrt{1-\omega_{\rm pe}^2/\omega_s^2}$). The initial peak amplitude of the seed equals to $0.05$. The pump is defined as a number ($N_{\rm pump}$) of identical Gaussian envelopes with the varied width and duration within $2-100$ range. The crossing angles range from $\pm 30^\circ$ to $\pm 180^\circ$, thus including both side and backward Raman amplification configurations. The pump group velocities, $v_{gp,i}$, are equal to each other and correspond to the group velocity of electromagnetic wave with $\omega_{p,i}=\omega_p$ travelling in plasma with plasma frequency $\omega_{\rm pe}$ ($v_{gp}= c\sqrt{1-\omega_{\rm pe}^2/\omega_p^2}$); pumps are synchronized with the seed in such a way that they intersect with the center of mass of the seed at $x=-125,~y=0$ at their own focal planes (in other terms, $5 z_R$ away from the seed focal plane); in case of multiple pumps in backwards Raman configuration, pumps intersect the seed at $y=\pm dy_i$, where $dy_i$ is the transverse shift of the laser axis of $i$-th pump. The amplitudes of the pump pulses defined from the seed to pump energy ratio, $r \equiv \mathcal{E}_{\rm seed,0}/\mathcal{E}_{\rm pump,0}$. As we are interested in nonlinear stage of Raman interaction to obtain better seed focusability \cite{Jia2020}, we fix $r=0.63$. We simulate the interaction for $\tau_{\rm fin} = 2000$, i.e. until the seed pulse reaches its focal plane. Figure \ref{fig:setup} shows a typical setup that we consider in NSE simulations.

\section{Simulation results}\label{results}
We consider various pump configurations with NSE simulations and interpret them using the results of Section \ref{theory}. 

\subsection{Narrow single pump pulse in backward Raman scattering}

\begin{figure}
    \centering
    \includegraphics[width=\linewidth]{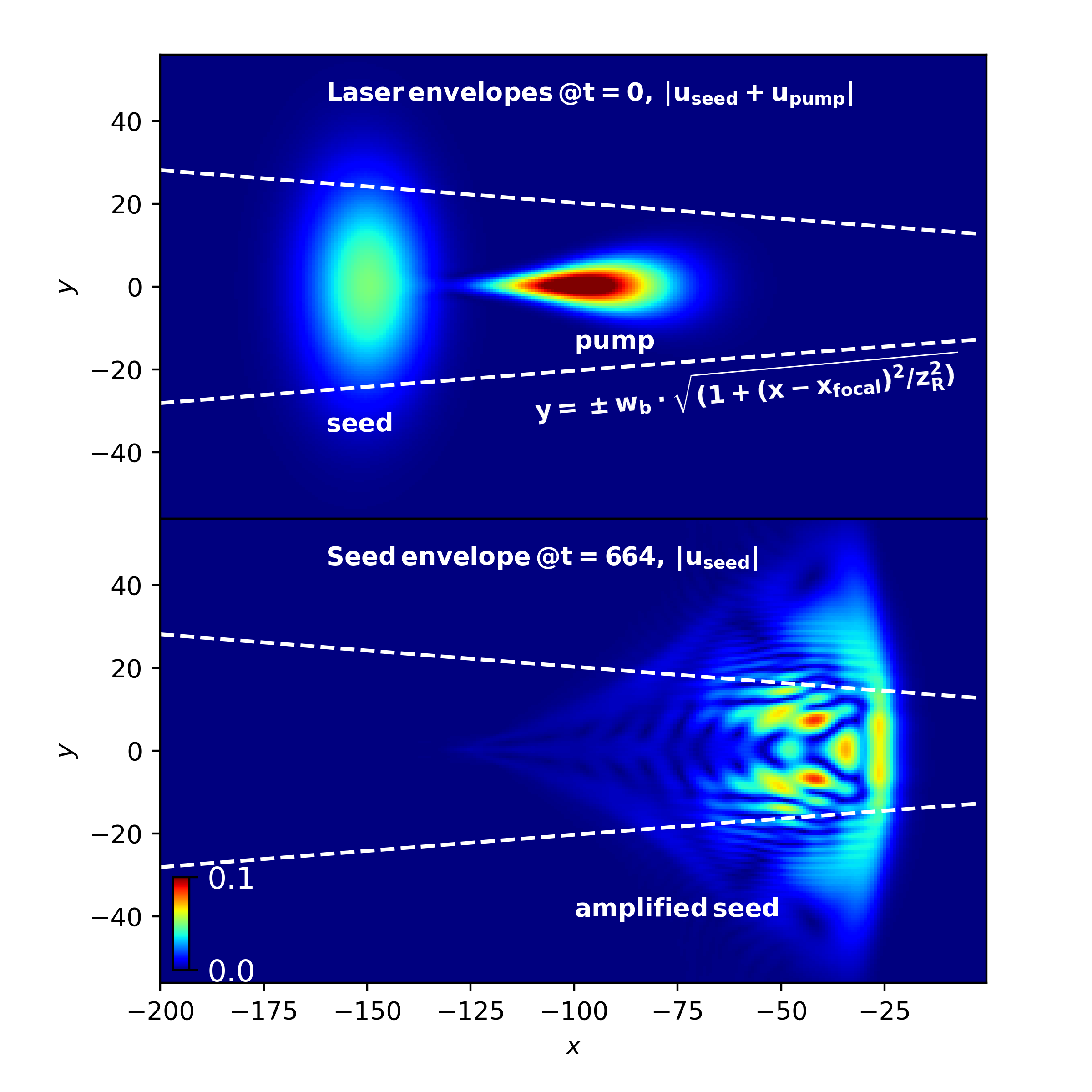}
    \caption{Intial and final stages of the NSE simulation of seed-single pump interaction in backward Raman scattering regime. Dashed lines depict the evolution of the Full Width Half Maximum of the original seed pulse.}
    \label{fig:case1}
\end{figure}

First, let us discuss the single pump-seed interaction in the backward Raman scattering regime. Figure~\ref{fig:case1} shows the initial and final states of seed and pump envelopes. Dashed lines depict the evolution of the Full Width Half Maximum of the original seed pulse. Pump pulse with width $w_a=2$ and duration $\tau_a=20$ loses a significant portion of its energy, $\approx 75 \%$. After the interaction we may see a rather distorted seed pulse envelope with lost focusability, as might be expected from theoretical considerations. Figure~\ref{fig:intensityvswidthvsz} shows the evolution of the intensity of the amplified seed along the pulse propagation, $I(r=0,z)/I_0$. Right after the amplification plane ($z=-5z_R$, vertical dashed line), $I(r=0,z)/I_0$ starts to drop for the case of $w_a=2$ all the way to focal plane, $z=0$. The filamentary structure of the seed in NSE simulations may be attributed to the combination of $\pi$-pulse-like structure known to appear in backward Raman amplification \cite{Malkin1998} and diffraction of narrow amplified regions of the seed.

\subsection{Wide single pump pulse in backward Raman scattering}
\begin{figure}
    \centering
    \includegraphics[width=\linewidth]{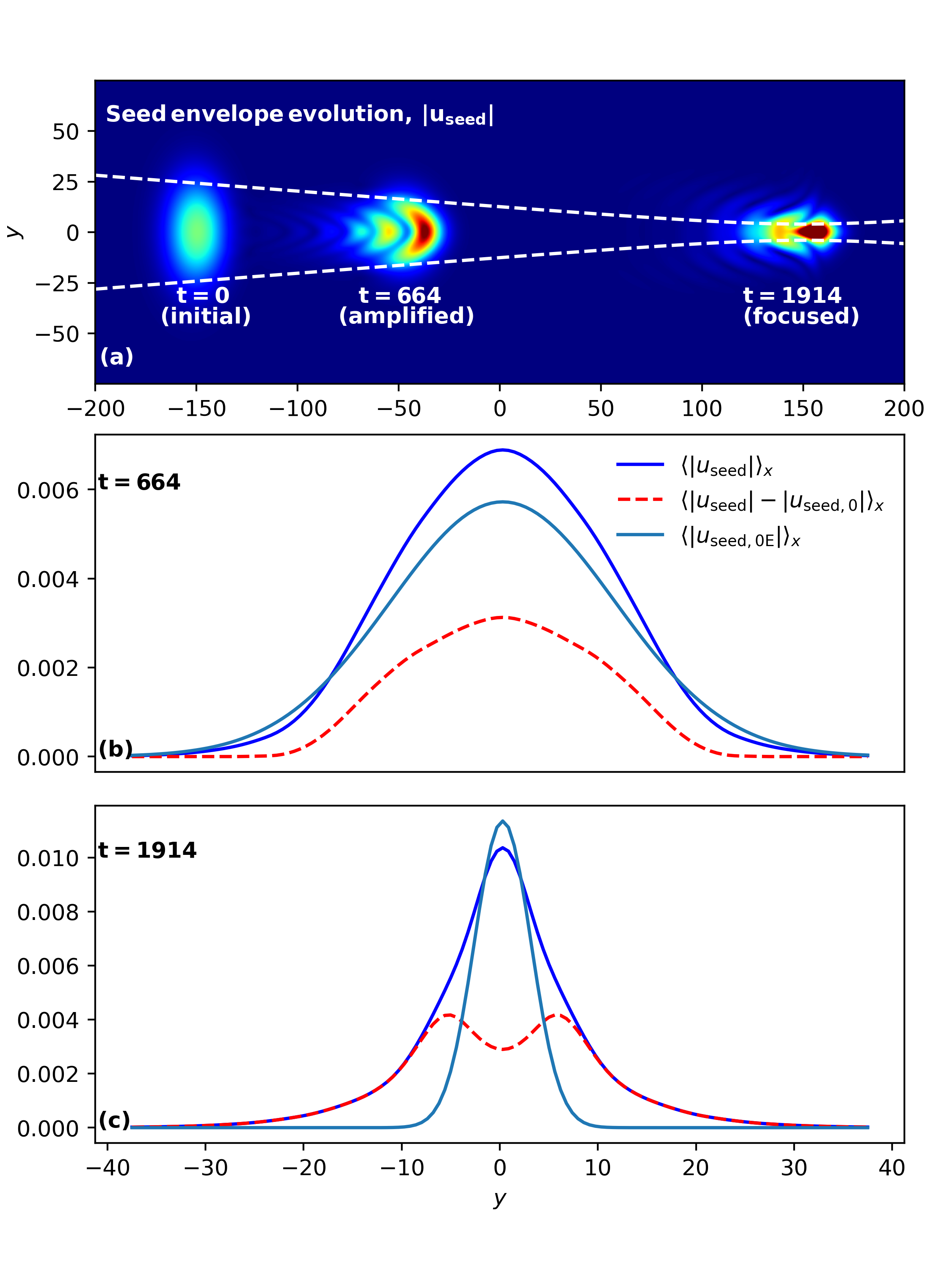}
    \caption{(a) Evolution of seed envelope in the single pump-seed interaction in backward geometry for $w_a=20$; three stages are shown: initial (t=0), right after the energy transfer is over (t=664), and around the original focal plane of the seed (t=1914). White lines are given by $y= w_b \cdot \sqrt{1+(x-x_{\rm focal})^2/z_R^2}$. Longitudinally averaged envelope profiles for t=664 (b) and t=1914 (c), showing amplified seed (blue solid line), amplified seed with subtracted unamplified seed envelope (red dashed line), and gaussian seed with the energy of the amplified seed (navy blue solid line).}
    \label{fig:case2}
\end{figure}

As predicted by analytical considerations, there exists an optimal pump width maximizing the focal plane intensity. Thus, we consider an increased pump width for the same total pump energy as in previous case. Figure \ref{fig:case2} shows the intial, intermediate (right after the energy transfer is over), and final stages of seed-single pump interaction in backward Raman scattering regime for the case of wider pump pulse, with $w_a = 20$. As we showed earlier, this value results into the optimal focal intensity of the amplified seed. We also calculate the longitudinally averaged transverse 1D profiles of the seed envelopes, $\langle |u_{\rm seed}| \rangle_x$ (blue solid lines), difference between longitudinally averaged 1D transverse profile of the amplified seed and analogous freely propagating Gaussian pulse, $\langle |u_{\rm seed}|-|u_{\rm seed,0}| \rangle_x$ (red dashed lines), and longitudinally averaged 1D transverse profile of the freely propagating Gaussian pulse with the same energy as amplified seed case (navy blue solid lines). First of these metrics highlights the average transverse profile of the amplified seed pulse; second metric estimates where in the tranverse profile the pump energy is deposited (thus, we subtract unamplified seed profile from amplified one); final metric is to compare the amplified seed with the Gaussian seed of elevated energy.

One may see that at the intermediate stage, $t=664$, longitudinally averaged peak intensity ($\langle |u_{\rm seed}| \rangle_x$, solid blue line), overcomes the one of the Gaussian envelope of the same energy ($\langle |u_{\rm seed,0E}| \rangle_x$, solid navy line). $\delta \langle |u_{\rm seed}| \rangle$ diagram shows bell shape with multiple changes of curvature sign. Later on, at the original focal plane of the seed ($t=1914$), one may see that the Gaussian envelope has a higher peak amplitude in comparison to amplified seed (Fig. \ref{fig:case2}c). The difference between amplified and non-amplified cases reveals two peaks of seed amplification displaced outside the $1/e$ region of the seed. The transient peak intensity reaching the Gaussian envelope level may be explained by positive interference of the amplified regions of the seed. Our analytical approach reaffirms this hypothesis, as may be seen in Figure \ref{fig:intensityvswidthvsz}. Orange and green lines ($w_a=4$ and 8, respectively) illustrate the transient intensity peak and intensity dropoff at the focal plane of the seed. From NSE simulation, we see that the seed width is effectively increased, and the additional energy is mainly deposited away from the laser axis, at $y\approx \pm 10$.

In the wide pump case one may also see that even though the longitudinal $\pi$-pulse-like structure is still there, the seed pulse experiences less diffraction due to wider amplified part of the seed, which, in its turn, diffracts slower according to Eqn.~\eqref{eqn:intensityonepump}.

\begin{figure}
    \centering
    \includegraphics[width=\linewidth]{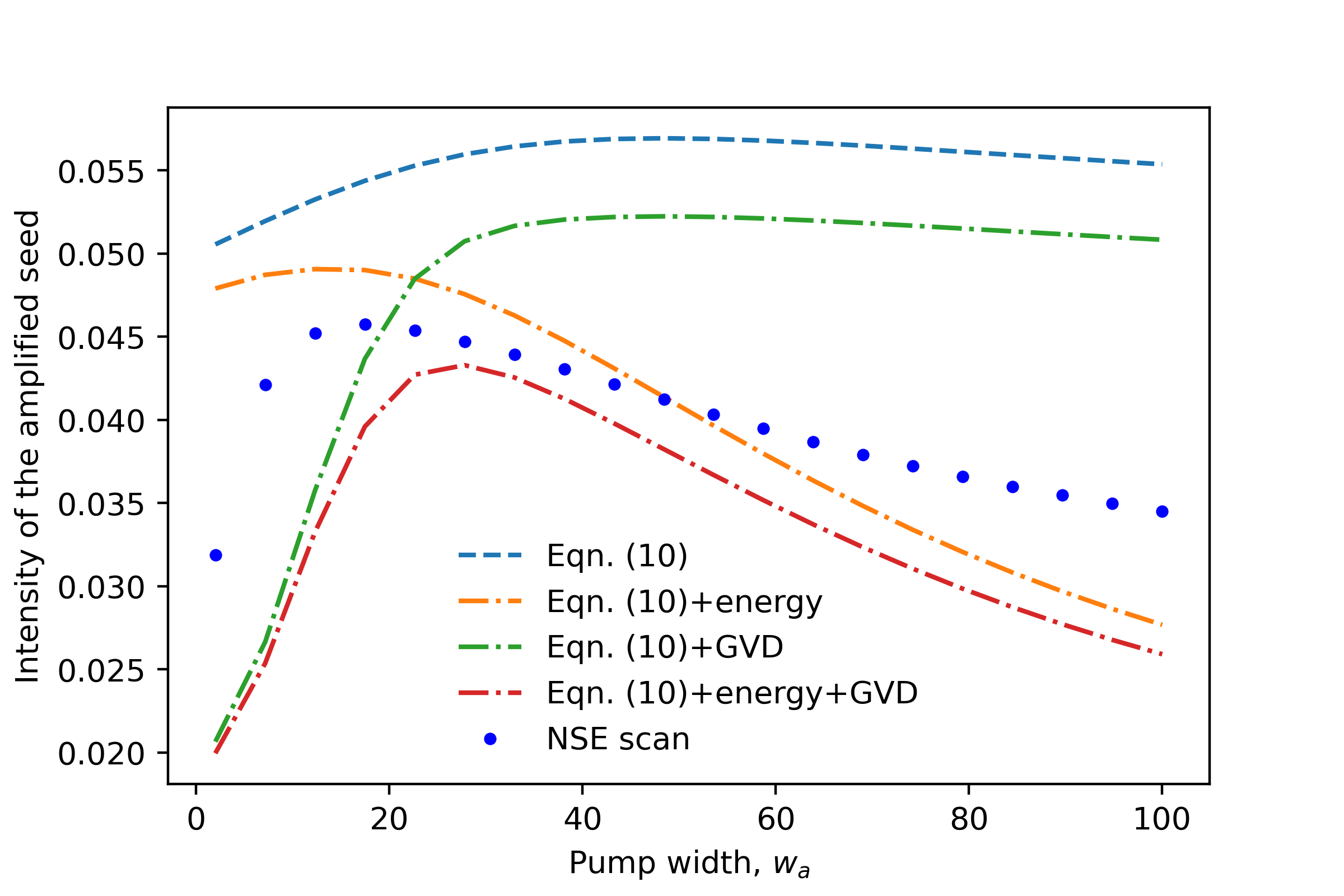}
    \caption{Comparison of the dependence of the peak amplified seed intensity at the seed focal plane as the function of pump width from analytical calculations and NSE simulations. Effects of focusing, incomplete energy transfer, and GVD are accounted for, and inclusion of all three leads to fair qualitative agreement with NSE simulations.}
    \label{fig:NSE_vs_theory}
\end{figure}

To make a more detailed comparison of the NSE simulations and analytical approach, we conduct a scan on pump width with NSE simulations and compare the peak focal intensity of the amplified part of the seed to the Eqn.~\eqref{eqn:intensityfocalplane}. By accounting for the effects of GVD and incomplete energy transfer into comparison between theory and NSE, we were able to obtain fair qualitative agreement with the NSE simulation results, see the result of such comparison for the typical NSE parameters and $w_a=2-100$ and Eqn.~\eqref{eqn:intensityfocalplane} evaluated for $z_0=-10 z_R$, $kz_R=2\pi z_R/\lambda$ on Figure~\ref{fig:NSE_vs_theory}.

\subsection{Two pump pulses in backward Raman scattering}
When increasing the pump width to improve the seed focusability is not possible, then, by using multiple pump pulses one may effectively recreate the required optimal transverse pump profile. As the next step in our simulations, we consider the backward Raman scattering in the case of two pumps of width $w_a=20$ each propagating in parallel with the transverse shift $dy=\pm 20$. The total pump energy is once again kept the same as in previous cases and may be calculated using Eqn.~\eqref{eqn:pumpenergy}.

For a relatively small pump separation, the two pump case will simply reproduce the single pump case with the same pump width. Then, by increasing the separation, we effectively create a plateau of pump intensity around the seed laser axis, which, as we shown using analytical considerations, is beneficial for focusability (see Eqn.~\eqref{eqn:twopumps} and Fig.~\ref{fig:Itwo2Ione}). Later on with the increase of pump separation, two Gaussian envelopes barely overlap, and the conditions for the positive interference of two pump pulses are violated.

\begin{figure}
    \centering
    \includegraphics[width=\linewidth]{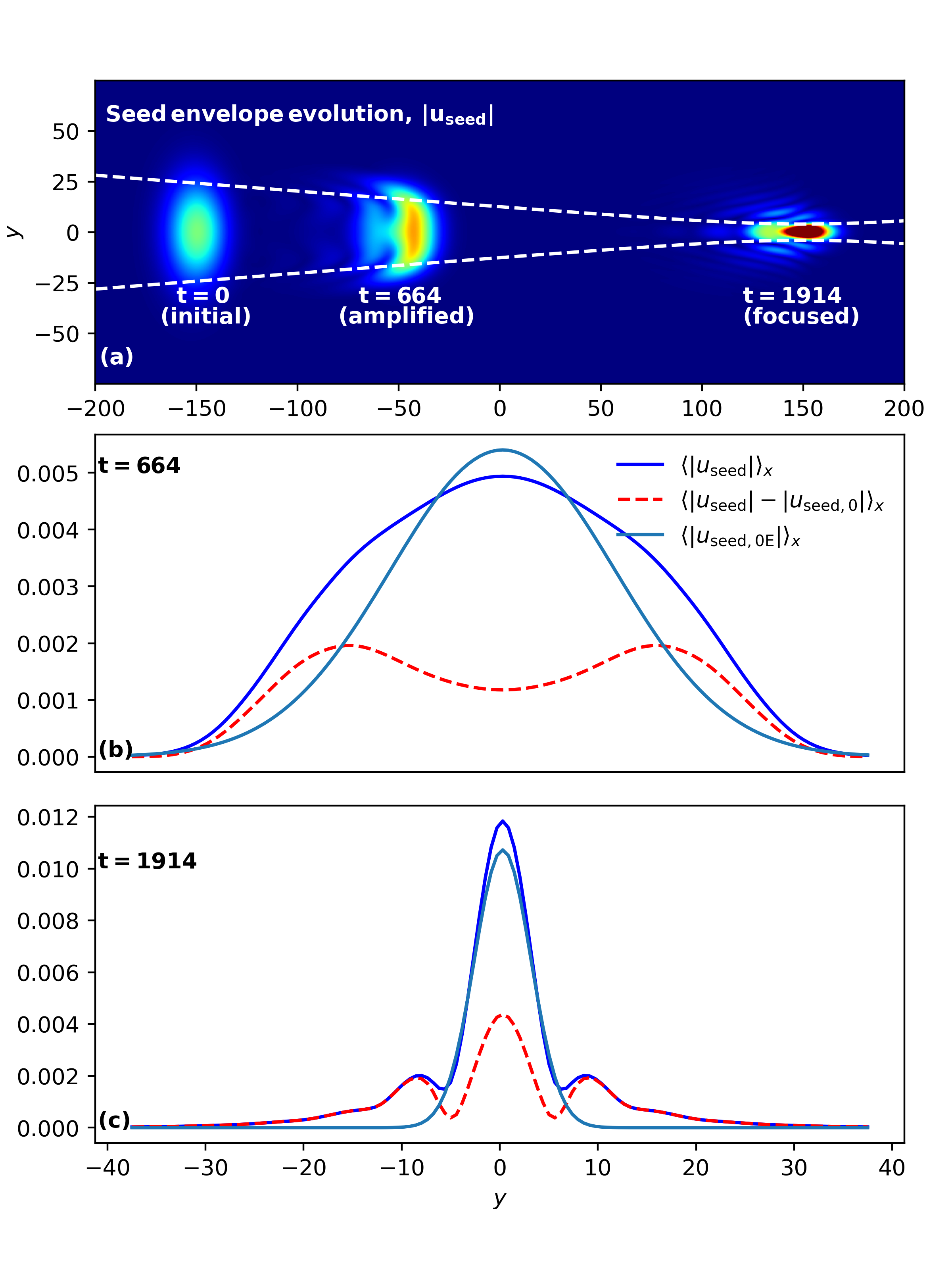}
    \caption{(a) Evolution of seed envelope in the two pump-seed interaction in backward geometry for $w_a=20$, $dy=20$; three stages are shown: initial (t=0), right after the energy transfer is over (t=664), and around the original focal plane of the seed (t=1914). White lines are given by $y= w_b \cdot \sqrt{1+(x-x_{\rm focal})^2/z_R^2}$. Longitudinally averaged envelope profiles for t=664 (b) and t=1914 (c), showing amplified seed (blue solid line), amplified seed with subtracted unamplified seed envelope (red dashed line), and gaussian seed with the energy of the amplified seed (navy blue solid line).}
    \label{fig:case3}
\end{figure}

This behavior is demonstrated using NSE simulations. Figure \ref{fig:case3} shows the evolution of seed envelope in the two pump backward Raman amplification regime. We first observe a wider and less intense profile of the amplified seed in comparison to the elevated Gaussian envelope (Fig.\ref{fig:case3}b, $t=664$), with main amplificaiton taking place, as expected, around pump pulse axes. Later on, however, the amplified seed overperforms the elevated Gaussian, as the amplified parts of the seed intersect to create positive interference sinc$^2$-like structure at the focal plane (Fig.\ref{fig:case3}c, $t=1914$), which is further stressed by $\langle |u_{\rm seed}| - |u_{\rm seed,0}| \rangle$ metric.

\begin{figure}
    \centering
    \includegraphics[width=\linewidth]{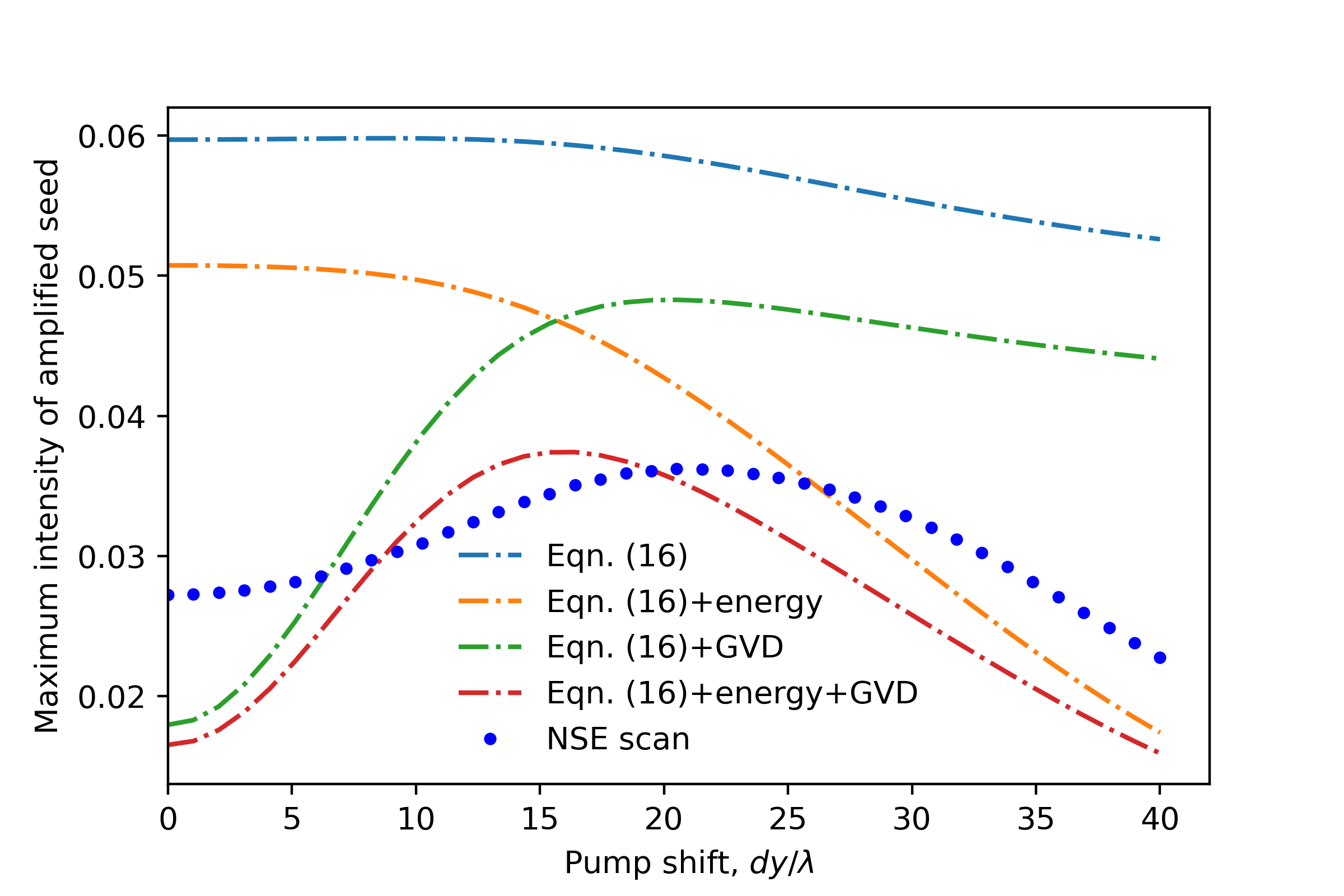}
    \caption{Peak intensity of the amplified seed as the function of pump separation from NSE simulations (blue circles) and theory (dot-dashed lines). Effects of focusing, incomplete energy transfer, and GVD are accounted for, and inclusion of all three leads to fair qualitative agreement with NSE simulations.}
    \label{fig:NSE_vs_theory2}
\end{figure}

As the analytical estimates predict, there is an optimal pump pulse separation that leads to the elevated intensity on the original focal plane, nearly reaching the theoretical limit. Conducting the NSE scan for the fixed pump parameters while only varying the pump separation, we obtained the dependence of the peak intensity of the amplified seed on the pump separation that may be compared to Eqn.~\eqref{eqn:twopumps}. Figure~\ref{fig:NSE_vs_theory2} represents the results of the NSE scan and Eqn.~\eqref{eqn:twopumps} evaluated for the parameters of the simulation. The qualitative agreement between NSE and simple theory is fair.

\subsection{One and two pump pulses in side Raman scattering}
\begin{figure}
    \centering
    \includegraphics[width=\linewidth]{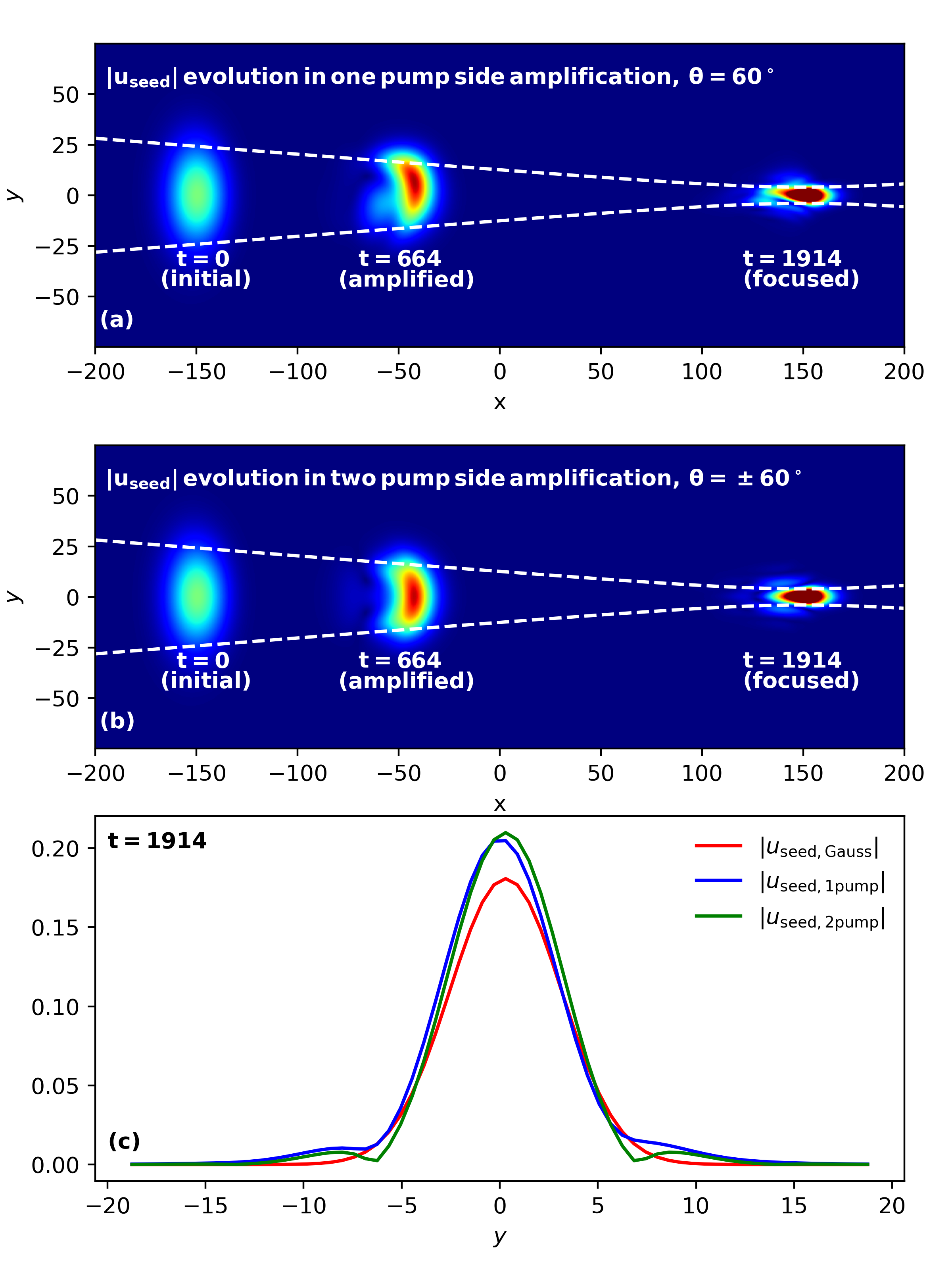}
    \caption{Evolution of seed envelope in (a) one and (b)two pump-seed interaction in oblique geometry for $w_a=20$, $\theta = \pm 60^\circ$; three stages are shown: initial (t=0), right after the energy transfer is over (t=664), and around the original focal plane of the seed (t=1914). White lines are given by $y= w_b \cdot \sqrt{1+(x-x_{\rm focal})^2/z_R^2}$. (c) Transverse profiles of elevated Gaussian seed (red line), single-pump (blue line) and two-pump (green line) amplified pulses.}
    \label{fig:case4}
\end{figure}

Another approach to improve the seed focusability is to evenly amplify the seed along the transverse direction, which may be realized via side Raman scattering geometry. In the non-linear Raman scattering regime the energy transfer is still efficient, on the level of backward Raman scattering, even though the coupling constant given by Eqn.~\eqref{eqn:NSEnorm3} drops with crossing angle $\theta$ as $\propto k_p^2+k_s^2-2k_pk_s\cos{(\theta)}$. At the same time, in the backward Raman scattering the amplification width, $w_a$, is identical to the pump width, whereas in the side scattering the effecive amplification width is around the seed width far away from the focal plane, $w_b$, with $w_b >w_a$. Figure \ref{fig:case4} illustrates side Raman scattering regime with one and two pumps crossing the seed with $\theta =\pm 60^\circ$ angles. The seed pulse gains around 68\% of the pump energy, and the {\it non-averaged} transverse profiles of the elevated Gaussian seed, single- and two-pump amplified pulses demonstrate remarkable coherence. The longitudinal modulations are avoided in the side scattering regime. It is also notable that the peak intensity of the amplified seed is observed at the original focal plane, in contrast to the previous cases, where we observed transient intensity peak which diffracted away at the focal plane. Our side Raman scattering NSE runs with one and two pumps show the elevated seed intensity on laser axis at the original focal plane, up to a factor of 1.3. The seed waist remains intact and the seed envelope shape is much more coherent in comparison to backward scattering cases.

\begin{figure}
    \centering
    \includegraphics[width=\linewidth]{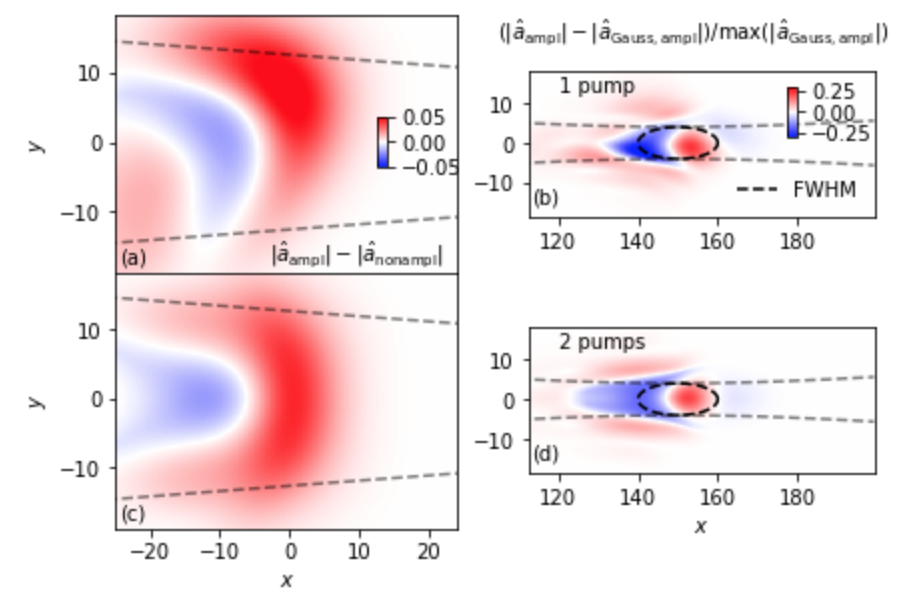}
    \caption{(a),(c) Spatial distribution of pump energy deposition after the seed-pump interaction and (b),(d) relative seed amplification by the seed-pump interaction compared to the free propagation of the Gaussian envelope of the same energy. Upper figures correspond to the strong seed regime with a single pump amplification, lower figures - 8 pump amplification with the one-point intersection arrangement. Grey dashed lines show the envelopes of the FWHM of the seed beam and black dashed ovals depict the FWHM of the seed at the focal plane.}
    \label{fig:2dprof}
\end{figure}

In case of side Raman scattering, the assumptions of the instantaneous and local amplification of the seed are indeed violated - the amplification happens along the trajectory of the pump pulse, which deposits energy at different planes within the seed pulse. Thus, one may expect that there will be effectively a continuum of the amplified parts of the seed interfering at the original focal plane that end up having an eleveated seed intensity. However, the match between these sources is not given. In order to understand what parts of the seed are actually amplified by the side Raman scattering, we consider two near-forward Raman scattering simulations with one and two pumps symmetrically amplifying the seed from angles $\pm 60^\circ$. Both runs have significant pump depletion efficiency, $\approx 67\%$. We calculate two two-dimensional distributions of products of the seed amplitudes: (a),(c) $|\hat{b}_{\rm ampl}|-|\hat{b}_{\rm nonampl}|$, which highlights the parts of the seed gaining additional energy in comparison to the identical non-amplified seed; (b),(d) relative seed amplification in comparison to the Gaussian envelope of the same energy as the amplified seed, $(|\hat{b}_{\rm ampl}|-|\hat{b}_{\rm Gauss,ampl}|)/|\hat{b}_{\rm Gauss,ampl}|$. Figure \ref{fig:2dprof} shows such diagrams, along with the boarders of the Full Width Half Maximum (FWHM) region of the original Gaussian pulse (dashed black lines), which are given by $y = \pm w_b \sqrt{1+(x-x_f)^2/z_R^2}$. It is evident that the majority of the pump energy is deposited within the FWHM region (Fig. \ref{fig:2dprof}a,c), with symmetric multi-pump arrangement improving the amplified seed symmetry. At the focal plane and within the FWHM ellipse, we also see an elevated seed amplitude, signifying the positive interference between the amplified parts of the seed (Fig. \ref{fig:2dprof}b,d). Both one- and two-pump configurations end up having the same intensity jump factor at the focal plane.

\section{Discussion}\label{discussion}

We investigated the focusability of the resonantly amplified seed pulse and its dependence on the pump configuration for a fixed total pump energy. Using coupled nonlinear Schroedinger equation (NSE) model, we conducted a comprehensive numerical scan, revealing the role of the pump width and, more generally, transverse profile of the seed amplification. 
It turns out that the uniform amplification of the seed within its FWHM region is optimal for preserving seed focusability, both in terms of a peak seed intensity at focal plane and preserving the amplified seed envelope. As demonstrated with NSE simulations, a uniform amplification profile may be obtained either using wide pumps (although at an expense of somewhat smaller pump depletion efficiency) or by setting up multiple Gaussian pumps in the transverse direction, which effectively create a plateau in pump amplitude, in the backward Raman scattering regime, or by considering side Raman scattering, which effectively resulted in a uniform amplification profile and, as a result, delivered an elevated intensity at focus (see Fig. \ref{fig:case4} and \ref{fig:2dprof}).

It should be noted that all simulations presented above are in the deeply nonlinear regime of Raman scattering, which is known to be the most effective in terms pump depletion \cite{Malkin1998,Trines2020} and preserving the original seed phase \cite{Jia2020}. Once we transition to the linear stage of Raman interaction, the seed focusability is no longer determined solely by the amplitude/intensity profile of the pump, and the pump optimization considerations presented in the paper do not apply.

Although we considered only ultrashort seed and pump pulses (i.e., pulse duration, $\tau_{\rm laser}$ is much smaller than typical ion response time, $\omega_{\rm pi}^{-1}$, $\omega_{\rm pi} \tau_{\rm laser} \ll 1$), for which the Raman scattering dominates Brillouin scattering mechanism, it may be that these results also apply to Brillouin amplification. Indeed, as the Brillouin scattering three-wave interaction is described by similar equations as Eqns.~\eqref{eqn:NSEnorm1}-\eqref{eqn:NSEnorm3} - see, e.g., Eqns. 4-6 in \cite{Lehmann2013} for 1D geometry and Eqns. 4-8 in \cite{Li2019} for 2D equations - one may expect similar focusability behavior as in Raman case. In \cite{Li2019}, effects of collisionality, thermal and ponderomotive self-focusing, and energy transport are also incorporated; however, if one considers a regime with $|a|^2,|b|^2 \ll 1$, $\nu_{ei}/\omega_0 n_e/n_{\rm cr} \ll 1$, non-relativistic ion acoustic velocity $c_s^2/c^2 \ll 1$, and nonlinear interaction regime, the Brillouin scattering equations end up having the same shape as Eqns.~\eqref{eqn:NSEnorm1}-\eqref{eqn:NSEnorm3}, up to the values of coupling constants $V$ and $W$. {It should be noted, however, that the energy transfer in the Brillouin case is generally slower than in Raman case due to lower growth rate, thus, the assumption of fast pump energy depletion is not necessarily satisfied there. Still, under the assumption of the strongly nonlinear regime of energy transfer between seed and pump pulse that was satisfied in some recent experiments (e.g., \cite{Marques2019}), our observations about focusability of Raman-amplified seed should apply to the Brillouin-amplified seed case as well.} In this respect, our results may suggest a way to improve the amplified seed focusability through using multiple pump beams available at NIF \cite{Kirkwood2018}.

The seed focusability in the backward Raman amplification case might be affected by self-focusing, as suggested by \cite{Li2018}. Here, we ignored self-focusing as we consider subcritical seed and pump pulse powers in terms of relativistic self-focusing \cite{Mori1997}. To double-check our qualitative understanding, we conducted auxilliary simulations with the self-focusing terms turned on (i.e. using Eqns.~\eqref{eqn:NSEpump} and \eqref{eqn:NSEseed} instead of Eqns.~\eqref{eqn:NSEnorm1} and \eqref{eqn:NSEnorm2}) showing only minor changes in the resulting metrics (seed intensity on focal plane, pump depletion, overall structure of the seed), which further verified the validity of our approach.

In summary, we have offered ways of controlling the transverse profile of the pump, which then serves to improve the seed focusability in the resonant amplification of the seed in plasma and may be of use for the design of the plasma-based amplification and beam combining.

This work was supported by NNSA DE-SC0021248.

\section*{Data availability statement}
The data that support the findings of this study are available from the corresponding author upon reasonable request.

\end{document}